\documentclass[aps,prl,showpacs,twocolumn,floats,superscriptaddress]{revtex4}

\usepackage{latexsym}
\usepackage[dvips]{graphics}

\begin{document}

\title{Simple optical measurement of the overlap and fidelity of quantum
states:  An~experiment}

\author{Martin Hendrych}

\affiliation{Joint Laboratory of Optics, Palack\'y University and
     the Physical Institute of the Czech Academy of Sciences,
     17.~listopadu 50, 772\,00 Olomouc, Czech~Republic}

\author{Miloslav Du\v{s}ek}
\author{Radim Filip}
\author{Jarom\'{\i}r Fiur\'{a}\v{s}ek}

\affiliation{Department of Optics, Palack\'y University,
     17.~listopadu 50, 772\,00 Olomouc, Czech~Republic}

\date{\today}

\begin{abstract}
We present the experimental results of measurements of the overlap of
both pure and mixed polarization states of photons.
The fidelity and purity of mixed states were also measured.
The experimental apparatus exploits the fact that a beam splitter can
distinguish the singlet Bell state from the other Bell states, i.e.,
it realizes projections into the symmetric and antisymmetric
subspaces of photons' Hilbert space.
\end{abstract}


\maketitle

Recently it was theoretically shown  by several authors that many important
characteristics of quantum states such as the purity, overlap,
and fidelity can be measured directly  \cite{radim,ek-hor} without
carrying out a complete quantum-state reconstruction.
However, the proposed experimental setups were rather complicated.
They involved cavity QED, non-linear optics, etc.
The only reported experimental test is based on NMR \cite{Fei02}.
In this Letter we show that for two qubits, encoded
into the polarization states of photons, the same goal can be achieved with a
simple beam splitter. Besides, the same setup with a beam splitter can
serve as the recently proposed \cite{FiDuFi} universal measurement device
programmed by a quantum state (where the polarization of one photon is
measured, while the polarization state of the other photon serves as the
``program'').

Let us consider a flip operator $ V$ in the Hilbert
space $\mathcal{H} \otimes \mathcal{H}$ of two distinguishable but
equivalent subsystems: $ V \, |\psi\rangle \otimes
|\phi\rangle = |\phi\rangle \otimes |\psi\rangle$. Let us further
consider a  factorable state $ \rho_{\mathrm A} \otimes
\rho_{\mathrm B}$ in the same Hilbert space. Then it
follows from a direct calculation that
${\mathrm{Tr}} ( V \rho_{\mathrm A} \otimes \rho_{\mathrm B} ) =
{\mathrm{Tr}} ( \rho_{\mathrm A} \rho_{\mathrm B}
)\equiv F$; note that $\rho_{\mathrm A} \rho_{\mathrm B}$ is
not a direct product. Taking into account the relation $ V =
\Pi^{+} - \Pi^{-}$, where $\Pi^{+}$
and $\Pi^{-}$ are projectors to the symmetric and
antisymmetric subspaces of $\mathcal{H} \otimes \mathcal{H}$,
respectively, one finally obtains
\begin{equation}
  {\mathrm{Tr}} \left( \rho_{\rm A} \rho_{\rm B} \right) =
  {\mathrm{Tr}} \left( \Pi^{+} \rho_{\rm A} \otimes
     \rho_{\rm B} \right) -
  {\mathrm{Tr}} \left(  \Pi^{-}  \rho_{\rm A} \otimes
      \rho_{\rm B} \right).
 \label{overlap}
\end{equation}
It means that if we were able to implement projections to the symmetric and
antisymmetric subspaces we could measure the overlap of two general
states living in $\mathcal{H}$. In particular, we would be able to
measure the overlap of two pure states, $|\langle \psi | \phi
\rangle|^2$, the fidelity $\langle \psi |  \rho | \psi \rangle$
comparing the state $\rho$ with the ``original'' state $| \psi
\rangle$, and also to calculate the Hilbert-Schmidt distance of two general
states:
\begin{equation}
d(\rho_{\mathrm A}, \rho_{\mathrm B}) =
  \left[ \frac{1}{2}
  {\mathrm{Tr}} \, (\rho_{\mathrm A}- \rho_{\mathrm B})^2 \right]^{1/2}.
\label{D}
\end{equation}

Besides, having a device realizing projective measurement
$\{ \Pi^{+}, \Pi^{-} \}$, we could experimentally implement the
simplest version of a quantum ``multimeter'' controlled by quantum
``software'' that has been proposed in Ref.~\cite{FiDuFi}. In this
concept, one qubit (the ``program'') determines the measurement basis for a
projective measurement on the other qubit. Of course, any such
measurement can be realized only approximately (with some error-rate). It
has been shown \cite{FiDuFi} that under given conditions, the optimal
multimeter that maximizes the average fidelity is represented by
a projective measurement on the ``program'' and ``data'' together. This
measurement is described exactly by projectors to the symmetric and
antisymmetric subspaces of the ``program-data'' Hilbert space.

\begin{figure}[!t!]
\centerline{\resizebox{0.8\hsize}{!}{\includegraphics*{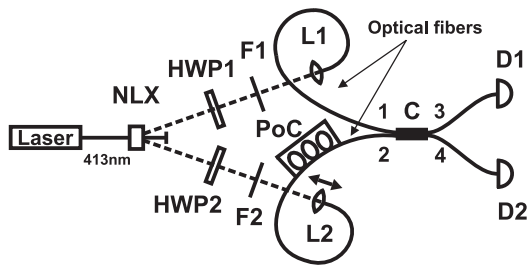}}}
  \caption{Setup of the experiment. NLX -- nonlinear
  crystal, HWP -- half-wave plates,  F -- long-wave pass filters (cut-off
  at 670\,nm), L -- lenses, PoC -- polarization controller, C -- fiber coupler,
  D -- detectors.}
  \label{scheme}
\end{figure}

Now let us turn our attention to the polarization states of two
photons. The states corresponding to the horizontal and
vertical linear polarizations will be denoted as $|{ H} \rangle$
and $|{ V} \rangle$, respectively. In such a case the projector into the
antisymmetric subspace has the following form $\Pi^{-} =
|\Psi^{-}\rangle \! \langle\Psi^{-}|$, where
\begin{equation}
  |\Psi^{-}\rangle = \frac{1}{\sqrt{2}} \Bigl(
  |{ H}\rangle_1 |{ V}\rangle_2 -
  |{ V}\rangle_1 |{ H}\rangle_2 \Bigr)
 \label{singlet}
\end{equation}
is nothing else but a singlet state \cite{BCJ}. What happens if a
singlet impinges on a beam splitter? It is an elementary exercise
to show that a beam splitter transforms it to the state
$(|{ H}\rangle_3 |{ V}\rangle_4 - |{ V}\rangle_3 |{H}\rangle_4 ) / \sqrt{2}$
(the labeling of the inputs and outputs
of the beam splitter is shown in Fig.~\ref{scheme}). Such a
state results in a simultaneous detection at both detectors
placed in modes 3 and 4
\cite{Weinfurter94,Braunstein95,innsbruck,Mattle96,Bouwmeester97}.
The singlet is the only one of the four Bell states (completing
the basis in the Hilbert space of two qubits) that produces such a
coincidence detection. The other Bell states make only one of the
detectors fire. This fact makes the measurement on two qubits
consisting of projectors $\Pi^{+}$ and $\Pi^{-}$
experimentally feasible. The beam splitter can be seen as a
universal quantum device suitable for the experimental realization
of all the tasks discussed above.

\begin{figure}[!t!]
\centerline{ \resizebox{0.82\hsize}{!}{\includegraphics*{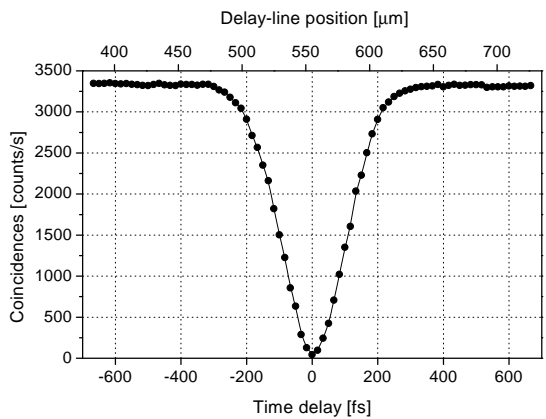}}}
  \caption{Hong-Ou-Mandel dip.}
  \label{dip}
\end{figure}

The experimental setup is shown in Fig.~\ref{scheme}. A
Kr$^{+}$-ion laser of wavelength 413.1~\,nm illuminates a
10-mm-long nonlinear crystal of LiIO$_{3}$ (cut
$\theta=90^{\circ}, \phi=0^{\circ}$), where spontaneous
parametric non-collinear degenerate type-I down-conversion occurs.
After passing through the respective half-wave plates (HWP), the
down-converted photons are coupled into optical fibers and
combined at a fused 50/50 fiber coupler, thus forming a
Hong-Ou-Mandel interferometer \cite{Mandel}. Since optical fiber
deforms polarization states, one of the arms of the interferometer
contains a polarization controller to match the polarizations of
signal and idler beams at the coupler.
The polarization controller consists of several loops of fiber acting as
a set of a half-wave plate and two quarter-wave plates.
Single-photon counting modules (employing silicon avalanche photodiodes
with quantum efficiency $\eta=51\%$) are placed at
the output ports of the coupler and electronics measure their
coincidence rate. With this setup, visibilities exceeding 98\,\%
were reached. Higher visibilities could not be reached due to the
fact that the splitting ratio of the fiber coupler was not exactly
50/50 and due to the imperfections of the half-wave plates. A
typical Hong-Ou-Mandel dip is shown in Fig.~\ref{dip}. Different
time delays were generated by moving the coupling lens L2 and the
tip of the fiber towards the nonlinear crystal. Moving the
coupling stage $200\,{\mu}$m away from the center of the dip also
served to measure the coincidence rate $C_{200}$ on the shoulder for
normalization purposes (it represents a half of the impinging-pair rate).
According to Eq. (\ref{overlap}) the overlap is calculated from measured
data as follows:
\begin{equation}
{\mathrm{Tr}}(\rho_A \rho_B) =1-2{\mathrm{Tr}}(\Pi^{-} \rho_A \otimes \rho_B)
=1-\frac{C_0}{C_{200}},
\end{equation}
where $C_0$ is a coincidence rate when the arms of the interferometer
are balanced.

Each data point at presented plots has been derived from 50--200
one-second measurement periods. On average, 3300 coincidences per
second were measured on the shoulder away from the dip.
Statistical errors are smaller than the symbols of points in the graphs.
Other errors stem from the non-unit visibility
and optical-path fluctuations. The accuracy of
polarization-angle settings was better than $\pm 0.3^\circ$.

\begin{figure}[!t!]
\centerline{\resizebox{0.82\hsize}{!}{\includegraphics*{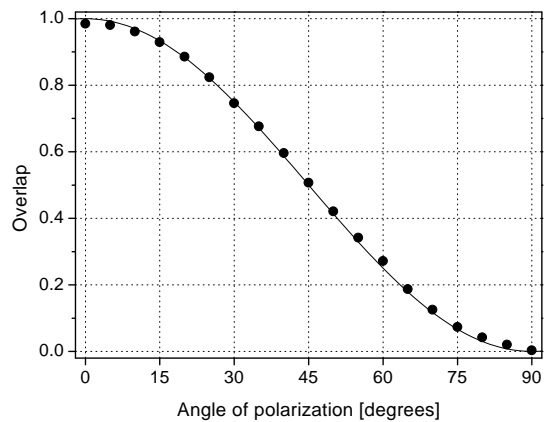}}}
 \caption{Overlap of two pure states. The first photon is in the state with
vertical linear polarization ($\theta = 0$). The other photon is also
linearly polarized but its polarization is rotated by angle $\theta >0$.}
  \label{overlapfig}
\end{figure}

The main results of this Letter are shown in Figs.~3--6. First we
measured the overlap of two pure states. Generated downconverted photon
pairs were linearly polarized in the vertical direction.
The polarization of photons in Arm 1 was kept fixed and the polarization
angle $\theta$  of photons in Arm 2 was varied by rotating the half-wave
plate HWP2. Thus we measured the overlap between the states $|V\rangle$
and $\cos\theta\, |V\rangle+\sin\theta\, |H\rangle$. The
experimental results shown in Fig.~3 are in very good agreement
with the theoretically expected dependence $\cos^2\theta$ that is
also plotted in Fig.~3.

\begin{figure}[!t!]
\centerline{\resizebox{0.82\hsize}{!}{\includegraphics*{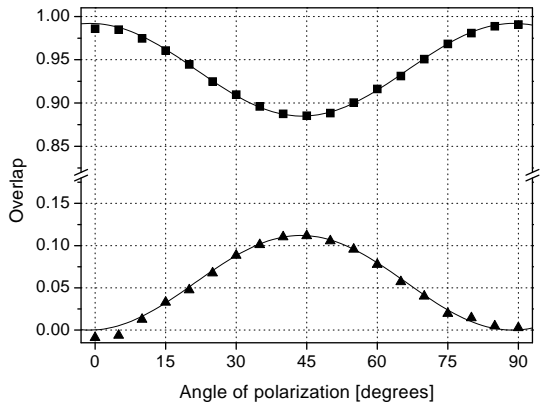}}}
 \caption{Overlap of two pure states (\ref{SISTATES}) with
 $\vartheta=\theta$ (squares) and $\vartheta=\theta+\pi/2$
 (triangles); $\vartheta$ is the polarization angle of the photon
 in Arm 1, $\vartheta$ is the polarization angle of the photon in Arm 2.}
  \label{baserot}
\end{figure}

In the next round of the measurements we simultaneously
changed the polarization states of both photons by rotating
both half-wave plates. In the first arrangement, both photons
were linearly polarized in the same direction $\theta$ after
passing through the half-wave plates and one would expect the
overlap to be equal to one for all $\theta$. In the second
arrangement the two photons were linearly polarized in
perpendicular directions, hence their polarization states were
orthogonal and the theory predicts $F=0$ for all $\theta$.

However, a different behavior was observed -- see Fig.~4. The
dependence of the overlap on angle $\theta$ can be explained by
the modification of the polarization state inside the fibers due to
birefringence and other effects. This effect must be
compensated for by the polarization controller whose proper setting
should ensure that if the two photons enter the fibers in the
identical polarization states, then they also arrive at the fiber
coupler in identical polarization states. It is relatively easy to
satisfy this condition for some chosen basis states, say
$|H\rangle$ and $|V\rangle$, for which the visibility of the dip
is tuned to maximum by manipulating the polarization
controllers; however, this does not guarantee that the above condition
will be satisfied for {\em any} polarization state. What happens
in the fibers is that the horizontally polarized photon acquires
certain non-zero phase shift with respect to the vertically
polarized one. This shift is different, in general, for the
fiber in Arm 1 and Arm 2. Therefore, the
polarization of the two photons at the coupler is not the same even if the
input polarization states are identical. However, this phase shift
does not play any role if at least one of the photons is in the
basis state $|H\rangle$ or $|V\rangle$. Due to the technical
difficulties with the fiber polarization controller, we were not
able to compensate for such phase shifts. The phenomenon can be
described by an effective phase shift $\phi$ of one of
the input polarization states. In our setup, we thus effectively
prepare the following polarization states of photons,
\begin{eqnarray}
|\psi_1\rangle &=& e^{i\phi}\cos\theta\, |V\rangle +\sin\theta\, |H\rangle ,
\nonumber \\
|\psi_2\rangle &=& \cos\vartheta\, |V\rangle +\sin\vartheta\, |H\rangle,
\label{SISTATES}
\end{eqnarray}
where $\theta$ and $\vartheta$ are controlled by rotating the
half-wave plates, and the fixed phase shift $\phi$ is a parameter
of our apparatus.  The outcomes of measurements with  $\vartheta=\theta$ and
$\vartheta=\theta+\pi/2$ are shown in Fig.~4.
The formulas for the overlaps read:
\begin{eqnarray}
F_{||}&=&1- \sin^{2}(2\theta) \sin^2(\phi/2),
\nonumber \\
F_{\perp}&=& \sin^{2}(2\theta) \sin^2(\phi/2).
\label{FIDELITIES}
\end{eqnarray}
The solid lines in Fig.~4 display the best fits of the form
$A+B\sin^2(2\theta)$, with $A_\perp=0$, $B_{\perp}=0.112$,
$A_{||}=0.992$, and $B_{||}=0.107$. A very good agreement between
the theory (\ref{FIDELITIES}) and experimental data is observed.
From the fit we can extract the absolute value of the phase shift,
$|\phi|= 39.4^\circ\pm0.3^\circ$. In the future experiment,
we plan to insert a Pockels
cell between the half-wave plate HWP1 and the fiber which will
allow us to compensate for the phase shift $\phi$ and vary it at will.

\begin{figure}[!t!]
\centerline{\resizebox{0.82\hsize}{!}{\includegraphics*{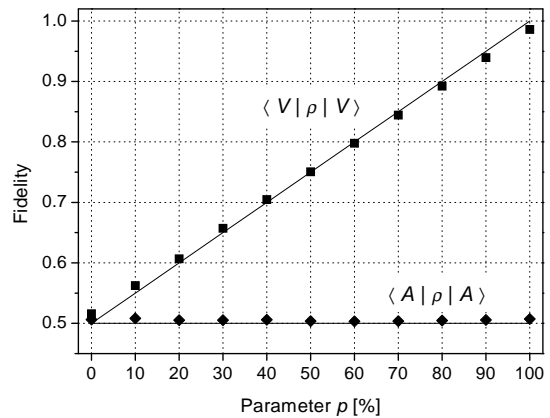}}}
 \caption{Measurement of the fidelity of a mixed state with respect to
 two different pure states $|V\rangle$ and $|A\rangle$.
 The full lines represent theoretical values.}
 \label{fidelity}
\end{figure}

So far we have focused on the overlap of two pure states. Our device can
also be used to measure the fidelity of a mixed state with respect to a
pure state or the overlap of two mixed states. The mixed state $\rho$
was created as a mixture of three pure states $|V\rangle$, $|X\rangle$
and $|Y\rangle$,
\begin{equation}
\rho = p \, | V \rangle \langle V | + \frac{1-p}{2} (| X \rangle \langle X |
+| Y \rangle \langle Y |),
\label{RHOMIX}
\end{equation}
where $| X \rangle$ and $| Y \rangle$ stand for two
mutually orthogonal polarization states. In the experiment, states
$|X\rangle$ and $|Y\rangle$ were generated by setting $\theta=45^\circ$
and $\theta=-45^\circ$, respectively.
The measured dependence of the fidelity $F= \langle\psi|\rho|\psi\rangle$
on the parameter $p$ is shown in Fig.~5 for two different
pure states $|V\rangle$ and $|A\rangle=(|V\rangle+|H\rangle)/\sqrt{2}$.

Let us now turn our attention to the overlap of two mixed states.
A particularly interesting case occurs if $\rho_{\rm A}=\rho_{\rm B}=\rho$
because the overlap  is then equal to the purity of $\rho$,
${\rm Tr}(\rho_{\rm A}\rho_{\rm B}) = {\rm Tr}(\rho^2)$.
To demonstrate this, we prepared both photons
in the same mixed state (\ref{RHOMIX}) following the procedure described
above. The measured purity  is plotted in Fig.~6 as a function of $p$.
If we know the purity $P$ of the qubit state $\rho$, we can
immediately determine the eigenvalues of $\rho$ from the formula
\begin{equation}
\lambda_{1,2}=\frac{1}{2}(1\pm \sqrt{2P-1}).
\label{LAMBDA}
\end{equation}
Our experimental setup thus enables a direct estimation of the
eigenvalues of $\rho$ without the necessity to reconstruct the
whole density matrix provided that two copies of $\rho$ are
available simultaneously for a joint measurement on $\rho\otimes
\rho$. The obtained eigenvalues are plotted in Fig.~6. They are in
good agreement with the theoretically expected behavior. The
largest errors of eigenvalues occur when $\rho$ is close to the
maximally mixed state where $ P\approx 1/2 $ and a small error in
$P$ causes a large error in $\lambda$ as can be deduced from
Eq.~(\ref{LAMBDA}). Note also that if we know the spectrum of
$\rho$, then we can determine several important characteristics of
$\rho$ such as the von Neumann entropy ${\rm
Tr}(\rho\ln\rho)=\sum_{j=1}^2 \lambda_j\ln \lambda_j$.

\begin{figure}[!t!]
\centerline{\resizebox{0.82\hsize}{!}{\includegraphics*{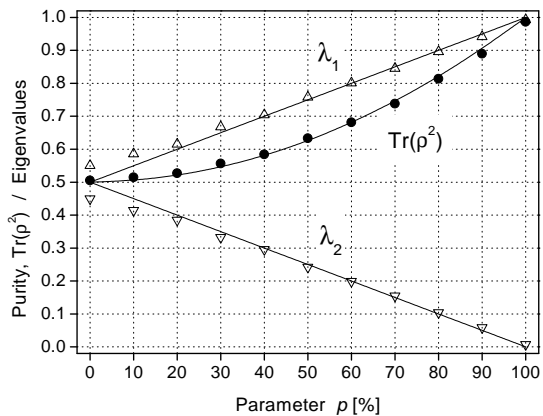}}}
\caption{Measurement of the purity of mixed state (\ref{RHOMIX})
-- circles. The full line represents the theoretical prediction
$P=(1+p^2)/2$. The triangles denote the eigenvalues of
corresponding density matrices, calculated from the purity.}
\label{purity}
\end{figure}

\begin{table}[!t!]
\caption{Measured overlap $F$ and Hilbert-Schmidt distance $d$ of
two mixed states. The columns labeled by $F_{\rm th}$ and $d_{\rm
th}$ show the theoretical values for comparison.}
\label{mix-tab}
\begin{ruledtabular}
\begin{tabular}{cccccc}
$p_{\rm A}$ &  $p_{\rm B}$ &  $F$&    $F_{\rm th}$ & $d$ &  $d_{\rm th}$ \\
\hline
0.2  &  0.4  &   0.545  &    0.540   &   0.097   &   0.100     \\
0.2  &  0.6  &   0.563  &    0.560   &   0.197   &   0.200     \\
0.2  &  0.8  &   0.581  &    0.580   &   0.298   &   0.300     \\
0.4  &  0.6  &   0.621  &    0.620   &   0.096   &   0.100     \\
0.4  &  0.8  &   0.658  &    0.660   &   0.201   &   0.200     \\
0.6  &  0.8  &   0.736  &    0.740   &   0.099   &   0.100
\end{tabular}
\end{ruledtabular}
\end{table}

Finally, we experimentally determined  the overlap of two
different mixed states $\rho_{\rm A}$ and $\rho_{\rm B}$ of the form
(\ref{RHOMIX}). The results for several different parameters $p_{\rm A}$
and $p_{\rm B}$ are summarized in Table~\ref{mix-tab}.
If we combine these data with
the direct measurement of the purity of states $\rho_{\rm A}$ and
$\rho_{\rm B}$, we can calculate the Hilbert-Schmidt distance
$d(\rho_{\rm A},\rho_{\rm B})$ from Eq.~(\ref{D}) -- the results are also
shown in Table~\ref{mix-tab}.

As mentioned in the introduction, our experimental device can also
serve as a ``quantum multimeter'' \cite{FiDuFi}. The polarization state
of one input photon, $|\psi\rangle$, represents a program (it
determines the measurement basis spanned by $|\psi\rangle$ and its
orthogonal counterpart $|\psi_{\perp}\rangle$). The other input
photon represents the measured qubit (in some ``unknown'' state
$|\varphi\rangle$). A coincident detection corresponds to the
measurement result ``one'', a detection at only one of the detectors
corresponds to the measurement result ``two''. The probabilities of the
results ``one'' and ``two'', respectively, read:
$p_{\perp,||}(\psi,\varphi)  =
\left[ 1 \mp |\langle \psi | \varphi \rangle |^2 \right]/2$. Of course, in
reality the performance of the multimeter is impaired by the low
detection efficiency. Nevertheless, we can still verify the
predicted fidelity of such a multimeter. It is given by the
formula \cite{FiDuFi}:
\begin{equation}
  F (\psi) = \frac{1}{2} \left[ p_{||} (\psi,\psi)
  + p_{\perp} (\psi,\psi_{\perp}) \right].
 \label{multim}
\end{equation}
As we can measure the overlaps, we can also determine this
function. For the three program states of the form $\cos\theta\,
|V\rangle +\sin\theta\, |H\rangle$, the corresponding fidelities
are shown in the following table:
\medskip
\begin{ruledtabular}
\begin{tabular}{cccc}
$\theta$: & $0^\circ$ & $-45^\circ$ & $45^\circ$ \\
$F$:  & 0.742 & 0.748 & 0.748
\end{tabular}
\end{ruledtabular}
\medskip
\noindent
The deviation from the theoretical value $3/4$ is mainly
due to the non-unit visibility.

\begin{acknowledgments}

This research was supported by the  Ministry
of Education of the Czech Republic, projects LN00A015 and RN19982003012.

\end{acknowledgments}


\begin{thebibliography}{99}


\bibitem{radim}
R.~Filip, Phys.\ Rev.\ A {\bf 65}, 062320 (2002).



\bibitem{ek-hor}
A.\,K.~Ekert {\em et al.}, Phys.\ Rev.\ Lett.\ \textbf{88}, 217901 (2002).


\bibitem{Fei02}
X.~Fei, D.~Jiangfeng, W.~Jihui, Z.~Xianyi, and H.~Rongdian,
quant-ph/0204049.


\bibitem{FiDuFi}
J.~Fiur\'{a}\v{s}ek, M.~Du\v{s}ek and R.~Filip, quant-ph/0202152.


\bibitem{BCJ}
S.\,M.~Barnett, A.~Chefles, and I.~Jex, quant-ph/0202087.


\bibitem{Weinfurter94}
H.~Weinfurter, Europhys. Lett. \textbf{25}, 559 (1994).


\bibitem{Braunstein95}
S.\,L.~Braunstein and A.~Mann, Phys.\ Rev.\ A \textbf{51}, R1727 (1995).


\bibitem{innsbruck}
M.~Michler, K.~Mattle, H.~Weinfurter, and A.~Zeilinger,
Phys.\ Rev.\ A \textbf{53}, R1209 (1996).


\bibitem{Mattle96}
K.~Mattle, H.~Weinfurter, P.\,G.~Kwiat, and A.~Zeilinger,
Phys.\ Rev.\ Lett.\ \textbf{76}, 4656 (1996);


\bibitem{Bouwmeester97}
D.~Bouwmeester, J.\,W.~Pan, K.~Mattle, M.~Eibl, H.~Weinfurter, and A.~Zeilinger,
Nature (London) \textbf{390}, 575 (1997).


\bibitem{Mandel}
C.\,K.~Hong, Z.\,Y.~Ou, and L.~Mandel,
Phys.\ Rev.\ Lett.\ \textbf{59}, 2044 (1987).



\end{thebibliography}
\end{document}